# Band Renormalization in Monolayer MoS$_2$ Induced by Multipole Screening

Woojoo Lee[1,2*], Seungwoo Yoo[3], Marios Zacharias[4], Junho Choi[3], Young-Kyun Kwon[3]

[1]Department of Physics, Gachon university, Seongnam 13120, South Korea
[2]Quantum Technology Institute, Korea Research Institute of Standards and Science, Daejeon, 34113, Republic of Korea
[3]Department of Physics, Kyung Hee University, Seoul, 02447, Republic of Korea
[4] Computation-based Science and Technology Research Center, The Cyprus Institute, Aglantzia 2121, Nicosia, Cyprus

Dielectric screening plays a crucial role in shaping the electronic structure of two-dimensional (2D) materials. In 2D semiconductors, screened Coulomb interactions arising from the surrounding dielectric environment are known to induce band renormalization, which is typically understood as a rigid shift of the electronic bands. Here, we experimentally demonstrate that dielectric screening can also give rise to non-rigid, momentum-dependent band renormalization. Using temperature-dependent angle-resolved photoemission spectroscopy (ARPES), we observe pronounced changes in the electronic band structure of monolayer MoS$_2$ on a highly oriented pyrolytic graphite (HOPG) substrate. The results indicate that temperature-driven variations in the effective interlayer separation modulate the dielectric screening experienced by monolayer MoS$_2$. At room temperature, the screening behavior is well described by a momentum-independent monopole approximation, whereas at liquid-helium temperatures the screening evolves into a multipole-like regime, leading to momentum-dependent band shifts.

## I. INTRODUCTION.

Two-dimensional (2D) van der Waals (vdW) materials have emerged as a versatile platform for realizing exotic quantum phenomena such as quantum spin, anomalous, and fractional hall effects [1–7], topological superconductivity [8,9], Moiré superlattice [10–14], and exciton condensation [15]. These phenomena are enabled by the exceptional tunability of vdW 2D systems, where stacking and twisting [16,17], and interlayer coupling provide unprecedented control over electronic structure and interactions [16,17]. At the same time, their low dimensionality makes them highly sensitive to the surrounding dielectric environment [18–21], which strongly modulates electron-electron interactions and thereby alters their intrinsic optical and electronic properties. Such sensitivity not only governs quasiparticle energies and excitonic responses [18–21] but can also induce nontrivial band renormalization that reshape the fundamental dispersions of 2D semiconductors [22]. A comprehensive understanding of dielectric screening is therefore critical for both uncovering emergent many-body phenomena and developing practical strategies for band-structure engineering in vdW heterostructures.

The endeavors to understand and harness the dielectric screening in 2D materials have been pursued primarily by investigating exciton physics [19,20,23,24]. The exciton is a bound state of electron and hole mediated via Coulomb interactions, which are screened by the surrounding dielectric environments. In 3D materials, the screened Coulomb interaction has the form of $1/\varepsilon_1\rho$, where $\varepsilon_1$ is a dielectric constant and $\rho$ is a distance between two charged particles. This form of Coulomb interaction retains the same dynamics as the unscreened case, $1/\varepsilon_0\rho$. However, in 2D materials, the screened Coulomb interactions deviates from its 3D counterpart depending on the length scale of $\rho$ and thickness ($h$) of 2D materials [25]. In the regime of $\rho \gg h$, Rytova-Keldysh (RK) potential has been widely employed to investigate excitons in 2D semiconductors [21,26–28] since the exciton radius extends to a few nanometers. On the other hand, RK potential does not provide a correct screened Coulomb interaction in the atomic scale regime [25,29,30].

Experimentally, Waldecker et al. [29] confirm the atomic scale regime of screened Coulomb interaction by directly observing electronic band structures on specifically located monolayer (ML) WS$_2$ on two different substrates; hBN and HOPG. Based on their study, the rigid band shift happens when the thickness of ML is much larger than the spatial extent of the orbitals in ML-WS$_2$. In a framework of image charge approximation, this condition results in monopole-like screened coulomb interaction, $\sim 1/\rho$. This approach is also understood with interlayer distance between ML and substrate. If the interlayer distance $d$ is far enough ($\rho \ll d$), the Coulomb interaction between a real charge in the ML and an image charge in the substrate is approximated to monopole-like form $\sim 1/\rho$. [30] However, when the interlayer distance decreases, higher order terms start to be involved (i.e., multipole interactions) [29].

*Corresponding author. Email: woojoolee@gachon.ac.kr

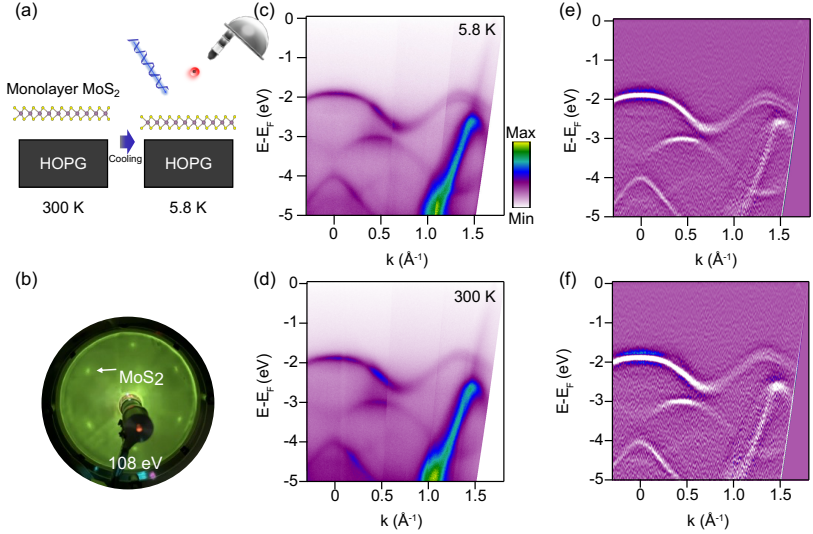

FIG. 1. Monolayer MoS$_2$ band structures acquired using ARPES. (a) Schematic of temperature-dependent ARPES experiment. (b) LEED image of the monolayer (ML) MoS$_2$ on a HOPG substrate. The six dots show the single-crystalline ML-MoS$_2$ structure. Additional characterizations on the ML-MoS$_2$ are provided in Supplementary Note 1 in Supplemental Material [31]. (c, d) ARPES Spectrum at 5.8 K and 300 K along the $\Gamma$- to $K$-points, respectively. (d, f) Second derivative images for enhancing the visibility of the bands in (c) and (d). The linear dispersions observed between 1 Å$^{-1}$ and 1.7 Å$^{-1}$ correspond to a Dirac band from the HOPG substrate.

The multipole screening regime has been rarely investigated compared to the monopole-like screening regime. Possibly because major optical phenomena occur near the direct band gap, where strong optical transitions take place. As a result, there has been less focus on the electronic states away from this region in energy-momentum space. Nevertheless, to fully harness the electronic properties of 2D materials, multipole screening effect is crucial since it modifies the electronic band structure non-monotonically, providing promising opportunities for engineering or stabilizing electronic structures in a desired way.

In this letter, we investigate the multipole-like screening effects on the electronic structure of ML-MoS$_2$ on a HOPG substrate. The interlayer distance between ML-MoS$_2$ and HOPG substrate is modulated by temperature control. At low temperature (5.8 K), ML-MoS$_2$ gets closer to the substrate, yielding a band hybridization between ML-MoS$_2$ bands with p$_z$ orbitals and HOPG $\pi$-bands. Simultaneously, non-rigid band renormalization occurs due to the multipole-like screening effects; the $K$-point valence band maximum (VBM) shifts down $\sim 170$ meV but $\Gamma$-point VBM does not shift. At high temperature (300 K), the interlayer distance gets larger, leading to the absence of band hybridization between the bands originated from p$_z$ orbitals in ML-MoS$_2$ and HOPG. At the same time, the $K$-point VBM turns back to their intrinsic VBM position, recovering its intrinsic ML-MoS$_2$ band dispersion.

## II. METHODS

### A. ARPES measurements.

ARPES measurements were performed using a helium discharge lamp with He-I$\alpha$ radiation (h$\nu$ = 21.2 eV) and a Scienta Omicron DA30 hemispherical electron analyzer. The overall energy resolution was better than 5 meV, as determined from the Fermi edge of a polycrystalline gold reference. During helium lamp operation, the chamber pressure was maintained below $2 \times 10^{-10}$ Torr.

### B. Sample preparation.

Large-area ML-MoS$_2$ ($3 \times 3\ mm^2$) was obtained using a gold-mediated exfoliation method from bulk MoS$_2$ crystals (Supplemental Material [31], Supplementary

*Corresponding author. Email: woojoolee@gachon.ac.kr

Note 1). This technique has been widely adopted to obtain millimeter-scale MLs ($> 1\ mm^2$) from various 2D van der Waals (vdW) materials [32–39]. In this process, the gold layer adheres more strongly to the outermost $MoS_2$ layer than the interlayer vdW interaction between adjacent layers, enabling the exfoliation of large-area monolayers [33].

Despite its outstanding performance, the use of gold can introduce contamination because polymers and gold etchants are required during the transfer process, which can impede surface-sensitive measurements such as ARPES [35,36]. To minimize contamination, we trim unnecessary regions of the sample and selectively remove the gold layer using a localized droplet etching method rather than immersing the entire structure in etchant solution. This procedure significantly reduces residue deposition on the monolayer surface and shortens the etching time, thereby minimizing potential sample damage (Supplemental Material [31], Supplementary Note 1). After transfer onto HOPG, the PMMA support layer was removed using acetone, isopropanol (IPA) and deionized (DI) water to eliminate etchant and polymer residues. The sample was subsequently annealed at 300 °C in a ultrahigh vacuum chamber overnight to achieve an atomically clean surface [40,41]. The high crystallinity of the obtained $ML-MoS_2$ was confirmed by LEED measurements, which reveal well-defined sixfold symmetry corresponding to $ML-MoS_2$ (Fig. 1b). The outer ring in the LEED image arises from a polycrystalline nature of the HOPG substrate.

### III. RESULTS AND DISCUSSION

The high-quality $ML-MoS_2$ surface allows us to observe a sharp electronic band structure in photoemission measurements. The individual bands are well-identified throughout the Brillouin zone. Especially, at $K$-point ($\sim 1.3\ Å^{-1}$) the spin splitting bands are clearly observed, which is easily buried by noise signals [42,43], highlighting the high sample quality. Besides the $ML-MoS_2$ bands, the Dirac band from a HOPG substrate is also identified in the momentum range between $1\ Å^{-1}$ and $1.7\ Å^{-1}$. The overall features are in good agreement with previous studies [40,41,44].

We vary the temperature over a wide range, from 300 K to 5.8 K, and observe its influence on the electronic band structure (Fig. 2). Distinct modifications occur in the bands near the $\Gamma$- and $K$-points. As the temperature decreases from 300 K, the energy position of the spin-split bands at the $K$-point shifts downward around 170 meV at 5.8 K. In contrast, the VBM at the $\Gamma$-point shows minimal shift (< 25 meV) (Fig. 2d). This observation indicates that the temperature-induced band shift results in non-rigid band renormalization. A similar temperature-dependent behavior of the bands was reported previously using CVD-grown polycrystalline ML-MoS$_2$ showing a relatively smaller band shift of 30 meV at $K$-point [45]. However, the discussion in that work mainly focused on linewidth broadening and valley-state robustness against interface defects, and the origin of the observed band shift was not discussed in detail.

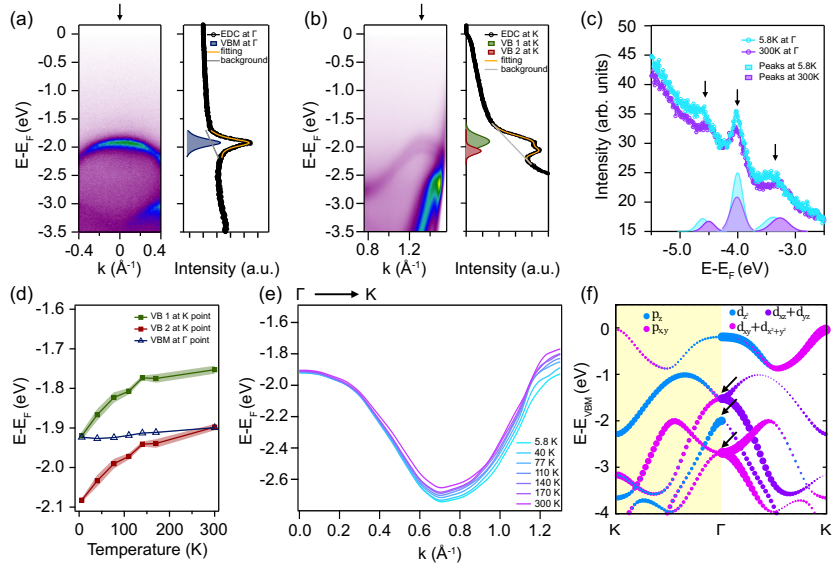

FIG. 2. Non-rigid band renormalization. (a, b) Energy distribution curves (EDCs) and ARPES spectrum images at $\Gamma$- and $K$-points, respectively. The black arrows above the spectra indicate the momentum positions for EDCs. (c) EDCs of the higher binding energy region at the $\Gamma$-point, taken at 5.8 K and 300 K. The peaks indicated by black arrows represent the band position at the $\Gamma$-point as shown in panel (f). (d) Temperature dependence of peak positions at $\Gamma$- and $K$-points. The triangle indicates VBM peak positions at the $\Gamma$-point, while the two squares (green and red) indicate peak points of the spin-split bands at the $K$-point. The raw data are provided in Fig. S11, Supplemental Material [31]. (e) Extracted band dispersion near the fermi-level, obtained by fitting. Near the $\Gamma$-point, the band remain unchanged but away from the $\Gamma$-point, deviation begin to appear. (f) Orbital-projected electronic band structure of monolayer $MoS_2$ from DFT calculations.

*Corresponding author. Email: woojoolee@gachon.ac.kr

Importantly, the influence of temperature variations on the electronic structure is reproducible even after multiple thermal cycles, ruling out the possibility of experimental artifacts (Supplemental Material [31], Fig. S7)

Specifically, the temperature-induced band renormalization predominantly occurs in bands composed of in-plane sulfur $p$ and molybdenum $d$ orbital components. For example, the VBM at the $\Gamma$- and $K$-points are mostly composed of out-of-plane $d_{z^2}$ and in-plane $d_{x^2+y^2}$ orbitals. The distinct band behavior under temperature variation is also observed in the high binding energy region at the $\Gamma$-point. The energy distribution curves (EDCs) in Fig. 2c illustrate band shifts at the $\Gamma$-point at two different temperatures: 5.8 K and 300 K. The band positions, marked by black arrows, correspond to the calculated bands in Fig. 2f, respectively. As observed, the bands composed of in-plane orbitals ($d_{xy}$, $d_{x^2+y^2}$, $p_x$, $p_y$) exhibit shifts, whereas those composed of out-of-plane ($p_z$, $d_{z^2}$) remain almost unchanged. These orbital selective shifts are possibly caused by orbital-dependent screened Coulomb interactions [46]. Importantly, the relative and orbital-dependent peak shift cannot be produced by experimental artifacts such as sample drift, energy referencing errors, or overall thermal broadening, which would affect all bands in a similar manner. The observation of contrasting temperature responses among different orbitals therefore provides an internal consistency check and reinforces the intrinsic nature of the observed band renormalization.

The overall orbital-dependent behavior can be understood in light of previous GW calculations [47] for monolayer MoS$_2$, which showed that the dominant self-energy contributions are nonlocal and strongly orbital dependent. In particular, the self-energy correction associated with the $d_{z^2}/d_{z^2}$ hopping channel is largely local and mainly produces rigid shifts, whereas the in-plane $d_{xy}$ and $d_{x^2-y^2}$ orbitals exhibit pronounced nonlocal self-energy contributions that lead to momentum-dependent renormalization. This difference arises from the orbital geometry: the in-plane Mo $d_{xy}$ and $d_{x^2-y^2}$ orbitals strongly overlap with neighboring atoms and form an extended hopping network in the lattice plane, while the $d_{z^2}$ orbital has weaker in-plane overlap. Since nonlocal Coulomb exchange primarily renormalizes intersite hopping, bands derived from the in-plane orbitals are expected to exhibit larger shifts than those originating from the out-of-plane $d_{z^2}$ orbital [47]. Such behavior is consistent with a multipole dielectric screening mechanism acting on the nonlocal electronic structure, which we will discuss below. In this work we focus on the experimentally observed non-rigid band renormalization and leave a comprehensive microscopic treatment for future studies.

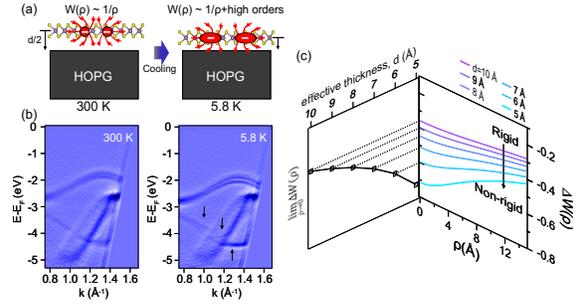

FIG. 3. Effects of interlayer distance change between ML-MoS$_2$ and HOPG. (a) Schematic of interlayer distance change caused by temperature variations. (b) Band hybridization effects at low temperature (5.8 K) caused by interlayer distance changes. The three black arrows indicate the modified ML-MoS$_2$ bands due to the hybridization at 5.8 K. (c) Calculated screened Coulomb interaction $\Delta W(\rho)$ using the analytical model.

We attribute the dramatic band shifts (Fig. 2e) to changes in the interlayer distance between the ML-MoS$_2$ and the HOPG substrate during temperature variation. We remark that the ML-MoS$_2$ heights, modulated by interlayer distance, can be varied from 6.6 Å to 10 Å depending on interface conditions (Supplemental Material [31], Fig. S6). The altered interlayer-distance modifies two primary factors affecting band renormalization: electronic hybridization [45] and dielectric screening. According to our density functional theory (DFT) calculations, which do not include dielectric Coulomb screening, electronic hybridization accounts for ~ 40 meV band changes for a 1 Å variation in interlayer distance (Supplemental Material [31], Fig. S8). The result is much smaller than the observed energy shift of ~ 170 meV, indicating that band hybridization alone cannot account for the observed effect. In contrast, dielectric Coulomb screening can shift the band by ~ 75 meV, based on calculations using the analytical model of screened Coulomb interaction $W(\rho)$, as shown below [30]:

$$W(\rho) = \frac{e^2}{\varepsilon_1 \rho} + 2\sum_{n=1}^{\infty} \frac{e^2 L_{12}^n L_{13}^n}{\varepsilon_1 \sqrt{\rho^2 + (2nd)^2}} + (L_{12} + L_{13})\sum_{n=0}^{\infty} \frac{e^2 L_{12}^n L_{13}^n}{\varepsilon_1 \sqrt{\rho^2 + [(2n+1)d]^2}} \quad (1)$$

where $L_{1n} = (\varepsilon_1 - \varepsilon_n)/(\varepsilon_1 + \varepsilon_n)$. $\varepsilon_1$, $\varepsilon_2$, $\varepsilon_3$ are dielectric constants of ML-MoS$_2$, vacuum and HOPG [29,48,49], respectively. The dielectric environment induced energy change, $\Delta W(\rho)$ is then calculated by subtracting $W(\rho, \varepsilon_1 = 10, \varepsilon_2 = 1, \varepsilon_3 = 9)$ from $W(\rho, \varepsilon_1 = 10, \varepsilon_2 = 1, \varepsilon_3 = 1)$, corresponding to cases with and without substrate screening (Fig. 3c). Within this framework, the conduction and valence band corrections are approximated as $\delta E_{c/v} = \pm \frac{1}{2} \lim_{\rho \to 0} \delta W(\rho)$ [30,50], leading to valence band shift being inversely proportional to the interlayer-distance between ML-MoS$_2$ and the HOPG substrate, i.e., $\delta E_v \propto -1/d$. We note that the effective thickness (d) used here does not correspond to a direct structural measurements of the physical interlayer distance; rather, together with the dielectric constants, it serves as an effective parameter, capturing interlayer-distance-dependent screening effects, as incorporated in image-charge frameworks in previous works [30,49]. This relationship was experimentally validated by inserting hBN layers between Graphene and WS$_2$. [20] Similarly, we anticipate that as the interlayer distance between HOPG and ML-MoS$_2$ decreases, the VBM at the K-point will shift downward rapidly, which aligns well with our experimental results in Fig. 2d. We emphasize that the analytic model used in this work is a minimal continuum description intended to provide physical intuition for the observed non-rigid renormalization, rather than a full microscopic dielectric treatment.

*Corresponding author. Email: woojoolee@gachon.ac.kr

Experimentally, the interlayer-distance change is distinctly captured at the band crossing points between ML-MoS$_2$ and HOPG bands. At 5.8 K, the ML-MoS$_2$ bands are modified by hybridization with HOPG $\pi$-bands near a binding energy of -4 eV, primarily composed of $p_z$ orbitals, as indicated by three black arrows in Fig. 3b. In contrast, at 300 K the interlayer distance is large enough to suppress the band hybridization. Additionally, no interactions are observed near -2 eV between ML-MoS$_2$ spin-split bands and Dirac band, due to the limited overlap between their $d_{x^2+y^2}$ and $p_z$ orbital components, consistent with previous reports [51]. The absence of hybridization in this energy range confirms that hybridization-induced band shifts are negligible. Importantly, we rule out thermal broadening as the origin of the hybridization features at 5.8 K, since the intrinsic linewidth of the MoS$_2$ band (~170 meV) greatly exceeds the thermal contribution (~25 meV) and therefore remains nearly temperature-independent (Supplemental Material [31], Fig. S11c). Thus, the revealed interlayer-distance changes, inferred from the presence or absence of band hybridization, underscore the significant role of dielectric screening in driving the non-rigid band shifts. We note that direct structural measurements of the temperature-dependent MoS$_2$–substrate separation, such as temperature-dependent XRD, STM, or cross-sectional TEM, are not available in the present study. Although our interpretation is supported by hybridization trends and screened-Coulomb modeling, quantifying the interlayer spacing at low temperature remains an open experimental task for future work.

The observed non-rigid energy shift of ~170 meV at the K-point is comparable in magnitude to the rigid band shifts of ~180 meV seen for ML- MoS$_2$ interfaced with hBN and HOPG [29], indicating that the energy scale of our measurement is consistent with Coulomb-screening effects. Notably, previous GW calculations performed using the SternheimerGW method [22] have also predicted non-rigid band renormalization in monolayer MoS$_2$ where hBN and SiO$_2$ dielectric screening shifts the VBM downward at the K-point by ~100 meV, placing the theoretical and experimental renormalizations on the same order of magnitude.

Other effects, such as in-plane lattice changes and electron-phonon interactions (EPI) [52] in the ML-MoS$_2$, could also contribute to the band renormalization. However, we exclude these effects for causing a non-rigid band renormalization for the following reasons. First, we rule out the possibility of temperature-dependent in-plane lattice change in ML-MoS$_2$ by examining the momentum position of the band composed of $p_z$ orbital at -4 eV near the $\Gamma$-point (Supplemental Material [31], Fig. S9). Within the experimental errors, no significant temperature-dependent momentum changes were observed, indicating the absence of substantial in-plane lattice changes. To further verify this, we investigated the bulk MoS$_2$ to assess thermal effects on the lattice structure (Supplemental Material [31], Fig. S10). The results show that the electronic band structure of bulk MoS$_2$ remains unaffected by temperature variations, aside from rigid shifts caused by charging effects during the photoemission experiments. Consistent with this, DFT calculations for free-standing monolayer MoS$_2$ predict strain-induced band shifts opposite to those observed experimentally (Supplemental Material [31], Fig. S8b), indicating that strain cannot explain the observed band evolution. Second, we exclude the possibility of a large non-rigid band renormalization due to EPI. To demonstrate this, we performed electron-phonon calculations using the special displacement method [52] to evaluate the k-dependent band renormalization (i.e., the real part of the self-energy $\Sigma(k, \omega)$) within the adiabatic Allen-Heine framework [53]. Our results show that the valence band renormalization at the $K$- and $\Gamma$-points evolves with temperature in nearly the same way, indicating that EPI induces an approximately rigid band shift of these two states with temperature in ML-MoS$_2$ (Supplemental Material [31], Supplementary Note 2). While different phonon modes (in-plane and out-of-plane) couple to electronic states with different strengths, our k-resolved analysis indicates that these effects result primarily in uniform energy shifts for these states rather than momentum-dependent band distortions in the temperature range considered.

*Corresponding author. Email: woojoolee@gachon.ac.kr

## IV. CONCLUSIONS

Therefore, we infer that the observed non-rigid band renormalization is driven by the non-monotonic screening effect, accompanied by electronic hybridization effects. As previously highlighted by Waldecker et al., [29] when the thickness of ML becomes comparable to its spatial orbital extent, the screened Coulomb interaction $W(\rho)$ is no longer monotonic. Instead, it becomes momentum-dependent. In this regime, the Coulomb interaction can no longer be approximated by a simple $1/\rho$ dependence; instead, higher-order multipole contributions must be taken into account. In our case, this effect arises from the reduced interlayer distance, as illustrated in Fig. 3a and 3c. Consequently, $W(\rho)$ including higher order terms is responsible for inducing momentum-dependent band renormalization.

## ACKNOWLEDGMENTS

W.L. was supported by the Gachon University research fund of 2024(GCU-202500540001) and the National Research Foundation (NRF) funded by the Korean government (MSIT) (RS-2025-25443562). S.Y. and Y.K.K. are funded by NRF-2022R1A2C1005505 and NRF-RS-2024-00416976. J.C. was funded by NRF-RS-2024-00452558. We acknowledge that electron-phonon calculations were performed using computational resources from the EuroHPC Joint Undertaking and supercomputer LUMI [https://lumi-supercomputer.eu/], hosted by CSC (Finland) and the LUMI consortium through a EuroHPC Extreme Scale Access call.

*Corresponding author. Email: woojoolee@gachon.ac.kr

*Corresponding author. Email: woojoolee@gachon.ac.kr

*Corresponding author. Email: woojoolee@gachon.ac.kr

*Corresponding author. Email: woojoolee@gachon.ac.kr